\documentclass{aa}  
\usepackage[utf8]{inputenc}   
\usepackage[T1]{fontenc}
\usepackage{lmodern}
\usepackage{amsmath}
\usepackage{bm}               
\usepackage{natbib}           
\usepackage{graphicx}
\usepackage{txfonts}
\usepackage{lipsum}
\usepackage{subcaption}         
\usepackage{lscape}             
\usepackage{placeins}
\usepackage{float}

\begin{document}

   \title{The dynamically hazardous asteroid 2025 TV$_{10}$}
   \subtitle{A new co-orbital asteroid of Venus}
   \author{V. Carruba \inst{1,2}, R. Sfair\inst{3,1}, O. C. Winter\inst{1}.}
  \institute{S\~{a}o Paulo State University (UNESP), School of Engineering and Sciences, Guaratinguet\'{a}, SP, 12516-410, Brazil.
             \email{valerio.carruba@unesp.br}
             \and Laborat\'orio Interinstitucional de e-Astronomia, RJ 20765-000, Brazil.
              \and LIRA, Observatoire de Paris, Université PSL, Sorbonne Université, Université Paris Cité, CY Cergy Paris Université, CNRS,  92190 Meudon, France.}

   \date{Received November 16, 2025}
 
  \abstract
   {Twenty co-orbital asteroids of Venus are currently known, several of which may evolve into potentially hazardous asteroids (PHAs) over timescales of thousands of years.}
   {We report the identification and first dynamical characterization of 2025 TV$_{10}$, a newly discovered Venus co-orbital asteroid, and assess its potential collisional hazard to Earth.}
   {We performed numerical simulations of a large number of asteroid clones, and we studied their close encounters with Venus and Earth.}
   {The asteroid may leave its co-orbital configuration on timescales of 1200 yr. The orbit of 2025 TV$_{10}$ is one of the closest of the absolute minimum of the "minimum orbital intersection distance" (MOID) with Earth for the Venus co-orbital asteroids known to date.}
   {Owing to its orbital parameters, 2025 TV$_{10}$ represents one of the most dynamically hazardous members of the Venus co-orbital population identified to date. Its faint magnitude and restricted observability windows make future observations challenging but essential for constraining its orbit.}

   \keywords{Minor planets, asteroids: general--Minor planets, asteroids: individual: 2025 TV$_{10}$--Venus--planets and satellites: terrestrial planets
               }

   \maketitle
\nolinenumbers
\section{Introduction}
\label{sec: intro}

Twenty co-orbital asteroids of Venus are currently known \citep{pan2024attemptbuilddynamicalcatalog, Carruba2025_Icarus}. These objects are in a 1:1 mean-motion resonance with Venus.  This resonance is regulated by the time behavior of the resonant angle $\sigma = \lambda -{\lambda}_2$, where ${\lambda}$ is the mean longitude of the asteroid and ${\lambda}_2$ that of Venus, respectively. At low eccentricities and inclinations we can observe tadpole orbits (TL4 and TL5), for which the resonant angle oscillates around one of the two stable Lagrangian equilibrium points L4 and L5 at $\sigma = 60^{\circ}$ and $300^{\circ}$, and horseshoe (H) orbits, for which a smaller object (such as an asteroid) shares nearly the same orbit  as a planet around a larger body (such as the Sun), but appears to move in a horseshoe-shaped pattern relative to that planet.  Currently, there is no known Venus co-orbital in a horseshoe configuration.  At higher eccentricities and inclinations, new equilibrium points may appear near $\sigma = 0^{\circ}$, and, from the planet's point of view the asteroid orbit may appear to move in a quasi-satellite (QS) configuration.  Compound orbits, which are combinations of horseshoe and quasi-satellite (HQS) and tadpole and quasi-satellite (TQS) configurations, are also possible at high eccentricities and inclinations \citep{Namouni1999a}. Sticking orbits are defined as those that evolved on just one side of the resonance and for which the change in the object's semimajor axis was at least one-fifth of the resonance width \citep{pan2024attemptbuilddynamicalcatalog}. Finally, orbits outside the resonance are defined as passing or circulating.

Transitions between these types of orbits may happen naturally for Venus co-orbital asteroids, which are all transient \citep{2006Icar..185...29M}. Typically, they remain in their resonant configurations for a co-orbital cycle lasting $12,000\pm6,000$ yr before transitioning to circulating orbits \citep{Carruba2025_Icarus}.  Some Venus co-orbital asteroids may experience repeated close encounters with Earth \citep{2000Icar..144....1C, 2006Icar..185...29M}. Some may become potentially hazardous asteroids (PHAs), which are asteroids that have an absolute magnitude H of 22 or less, and that may come within 0.05 astronomical units (au) of Earth's orbit.\footnote{https://www.minorplanetcenter.net/iau/Unusual.html} Six co-orbital asteroids of Venus have been identified as having the potential to become PHAs on timescales of thousands of years \citep{Carruba2025_Icarus}.   Recently, \citet{Carruba2025_AandA} modeled the possible threat that possibly undetected known co-orbital asteroids of Venus may pose by obtaining contour maps of the minimum orbit intersection distance (MOID) for clones of asteroids in an $(e, i)$ grid, with $e$ being the asteroid eccentricity and $i$ its inclination.

\begin{figure*}
  \begin{minipage}[c]{0.45\textwidth}
    \centering \includegraphics[width=2.3in]{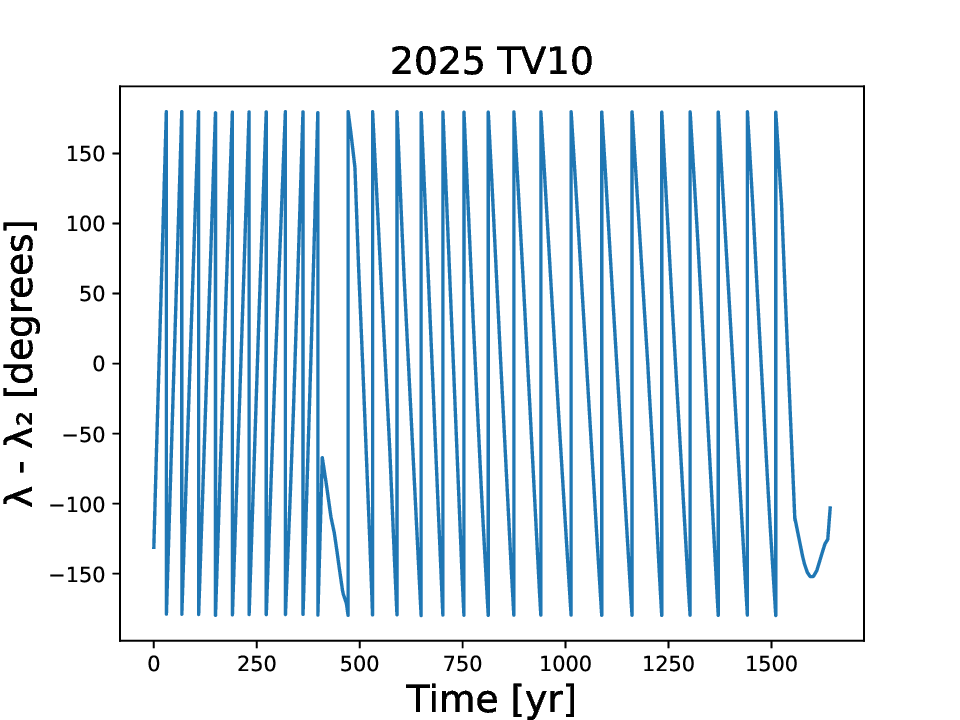}
  \end{minipage}%
  \begin{minipage}[c]{0.45\textwidth}
    \centering \includegraphics[width=2.3in]{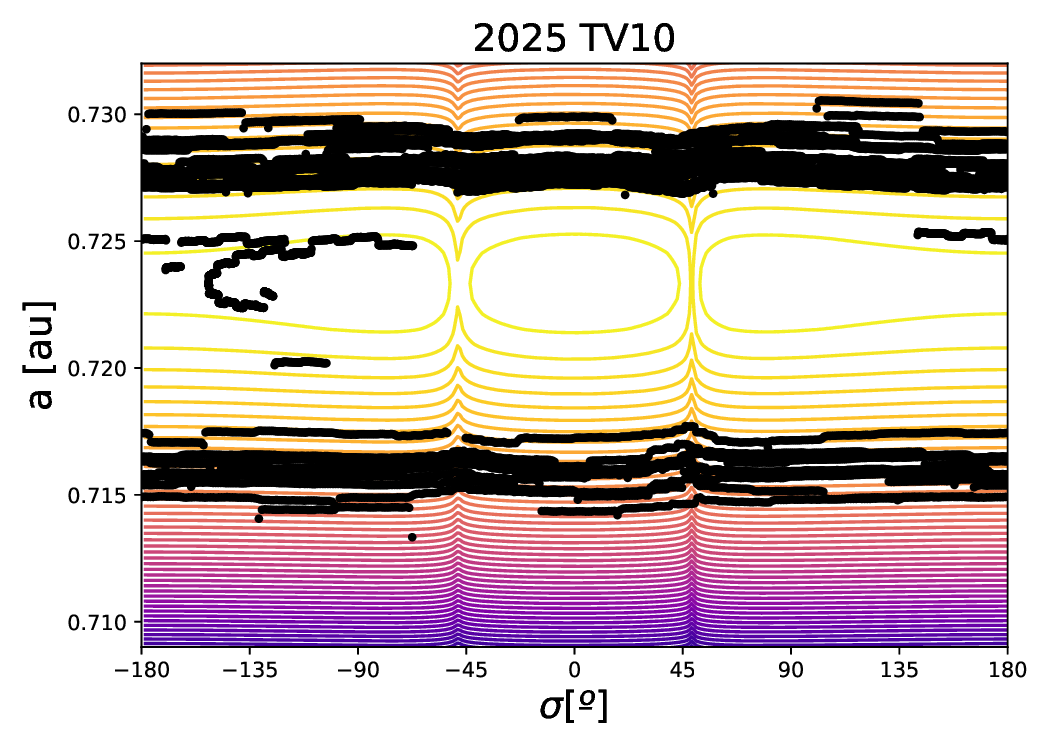}
  \end{minipage}
  \caption{Left panel: Time behavior of the resonant angle $\sigma = \lambda -{\lambda}_2$ for 2025 TV$_{10}$.  Right panel: Projections in the $(\sigma, a)$ plane of the Hamiltonian levels and the output of numerical simulations (black dots) for the same asteroid.}
  \label{Fig: res_config}
\end{figure*}

2025 TV$_{10}$ was first detected on October 15, 2025, by the Panoramic Survey Telescope and Rapid Response System 2 (Pan-STARRS 2), Haleakala \citep{Burgett2012PS2}.  It has an observation arc of 15 days, and an estimated absolute magnitude of 25.07.\footnote{See the JPL Horizons data available at: https://ssd.jpl.nasa.gov/tools/sbdb\_lookup.html\#/?sstr=2025\%20TV10 \\ The Minor Planet Electronic Circular (MPEC) is available at: https://www.minorplanetcenter.net/mpec/K25/K25U14.html\\ Finally, astrometry data can be retrieved from the MPC at:\\https://www.minorplanetcenter.net/db\_search/show\_object?\-utf8=\%E2\%9C\%93\&object\_id=2025+TV10} Its orbit is still poorly known, with an orbit condition code (OCC) of 7 (well-established orbits have OCCs of 0).  Currently, the MPC \citep{MPC} reports 58 observations from this asteroid.  A search performed using the ADAM Precovery service (B612 Asteroid Institute,\footnote{https://b612.ai/adam-platform/precovery/} \citep{Kiker2023ADAMPrecovery}) returned two valid precovery detections of (2025 TV$_{10}$): the first on December 10 2017 (MJD 58097.218), and the second on February 19 2021 (MJD 59264.220). We did not find any precovery detections with the EURONEAR Data Mining Tools\footnote{http://www.euronear.org/tools.php}, \citep{Vaduvescu2009EURONEAR}.  Attempts to compute an orbit including these two precoveries, however, were not successful, since the new measurements could not be fit with the rest of the dataset (Jon Giorgini, private communication).

Based on the 58 available observations from the MPC, the covariance matrix for 2025 TV$_{10}$, calculated at a MJD epoch of 2460964.5, corresponding to October 16, 2025 (Jon Giorgini, private communication), computed with the approach also described in \citet{Maggard2018Kuiper}, is shown in Table~\ref{Table: Covariance_matrix} in the Appendix. Table~\ref{Table: Nominal_orbit} in the Appendix shows the corrected elements at J2000 (Jon Giorgini, private communication). 

\section{Methods}
\label{sec: methods}

Numerical integration of the orbital elements of this asteroid, as reported in Table~\ref{Table: Nominal_orbit}, under the gravitational influence of the eight planets and the Moon with the SWIFT$\_$BS integrator, a Bulirsch-Stoer integrator of the SWIFT package \citep{2013ascl.soft03001L} with a tolerance factor of $10^{-8}$, shows that the object is first in a sticking orbit, transitions to a TL4 after $\simeq 350$ yr, and  transitions again to a sticking orbit after $\simeq 500$ yr, but this time at a larger semimajor axis than that of Venus (see Fig.~\ref{Fig: res_config}, left panel).  This is confirmed by the results of the semi-analytical model of \citet{pan2024attemptbuilddynamicalcatalog}, based on the model developed by \citet{2020CeMDA.132....9G}, which we show in Fig.~\ref{Fig: res_config}, right panel.  Because of its sticking orbital configuration, 2025 TV$_{10}$ is the 21st known co-orbital asteroid of Venus. 

With Lyapunov times of about 150 yr \citep{Tancredi1998}, co-orbital asteroids of Venus are extremely chaotic objects that can come into close contact with multiple planets. A single orbit may not carry much information about the real dynamical evolution of such asteroids, but statistical methods may provide valuable hints \citep{2000Icar..144....1C}. For this purpose, we first  generated 10000 clones of 2025 TV$_{10}$ to account for the orbital dispersion caused by the uncertainty in the orbital solution, according to the covariance matrix of the orbital elements. The covariance matrix was sampled by drawing multivariate normal random vectors whose mean equals the asteroid’s nominal orbital-element solution and whose $6 \times 6$ covariance equals the published uncertainty matrix in Table~\ref{Table: Covariance_matrix}. A Cholesky-based algorithm \citep{cholesky1924} transformed independent Gaussian deviates into correlated variations consistent with the covariance structure. Each draw yielded one statistically valid perturbation of the orbital elements, producing an ensemble of clones that reflects the uncertainty region in the six-dimensional element space.

\section{Results}
\label{sec: results}

The asteroid clones were integrated with the same scheme described in
Sect. 2, over 6000 yr, a time by which most clones should have left the co-orbital configuration.  We then tracked close encounters of the test particles with the Earth or Venus, which are detected when the particles enter a planet’s Hill radius ($R_H$) \citep{1991Icar...92..118H, 1992Icar...96...43H}:

\begin{equation}
R_H = a_p (1-e_P)  \sqrt[3]{\frac{M_p}{3(M_p+M_{\odot})}}.
\end{equation}

\begin{figure}
  \centering
  \includegraphics[width=3in]{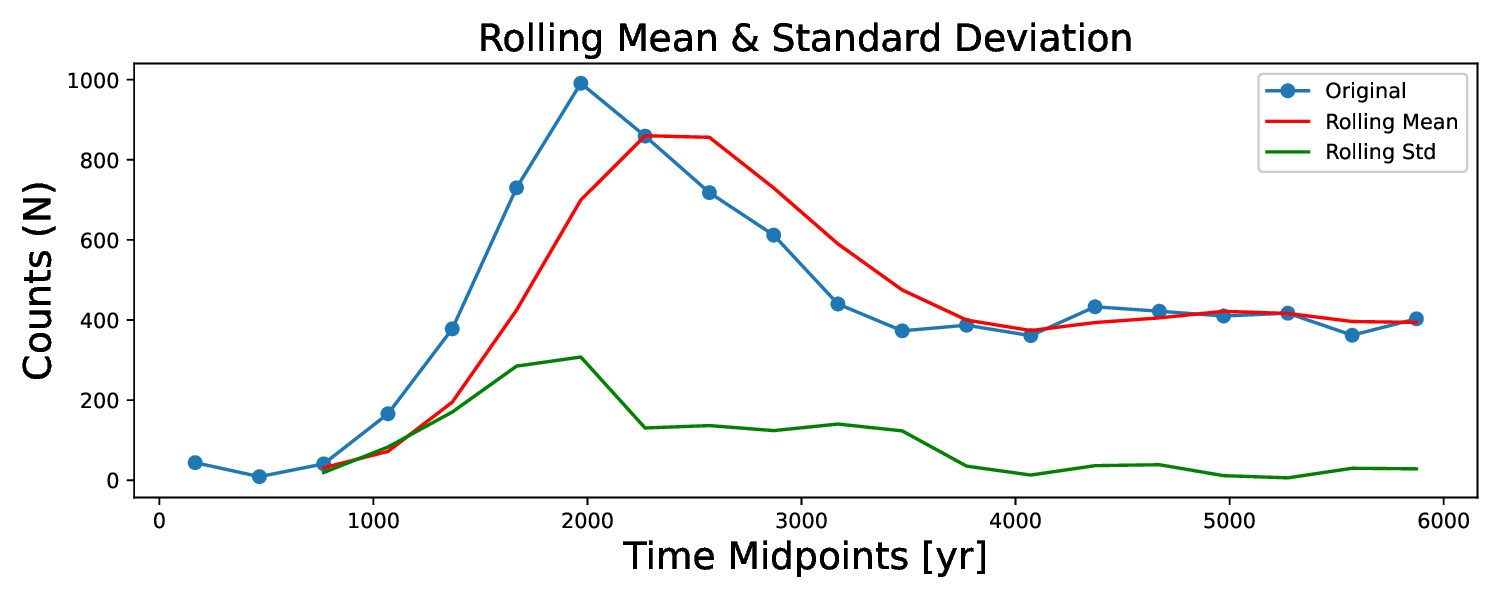}
  \caption{Time series of Venus encounters as a function of time (blue line).  The rolling standard deviation is shown by the green line, and the rolling mean is depicted by the red line.}
  \label{Fig: series_Venus}
\end{figure}

\noindent Here $a_p$ is the planet’s  semimajor axis, $e_p$ its eccentricity, $M_p$ its mass, and $M_{\odot}$ the Sun’s mass.  For Venus, $R_H$ is 0.0067 au, while for Earth it is 0.0098 au.  If a clone is in a co-orbital state with low amplitude, it is protected from close encounters with its perturbing planet.  Figure~\ref{Fig: series_Venus} displayed a time series of close encounters of the 10000 clones of 2025 TV$_{10}$ with Venus. We computed a rolling average of the mean and standard deviation of the number of encounters. Tests for detecting the stationarity of the time series were performed according to the methods described in \citet{Carruba2025_Icarus}. The asteroid remains in a co-orbital configuration for about 1200 yr, and after 3500 yr the time series of close encounters with Venus becomes stationary, suggesting that by then the majority of the asteroid clones are no longer in a co-orbital configuration.

\begin{figure*}
  \begin{minipage}[c]{0.45\textwidth}
    \centering \includegraphics[width=3.5in]{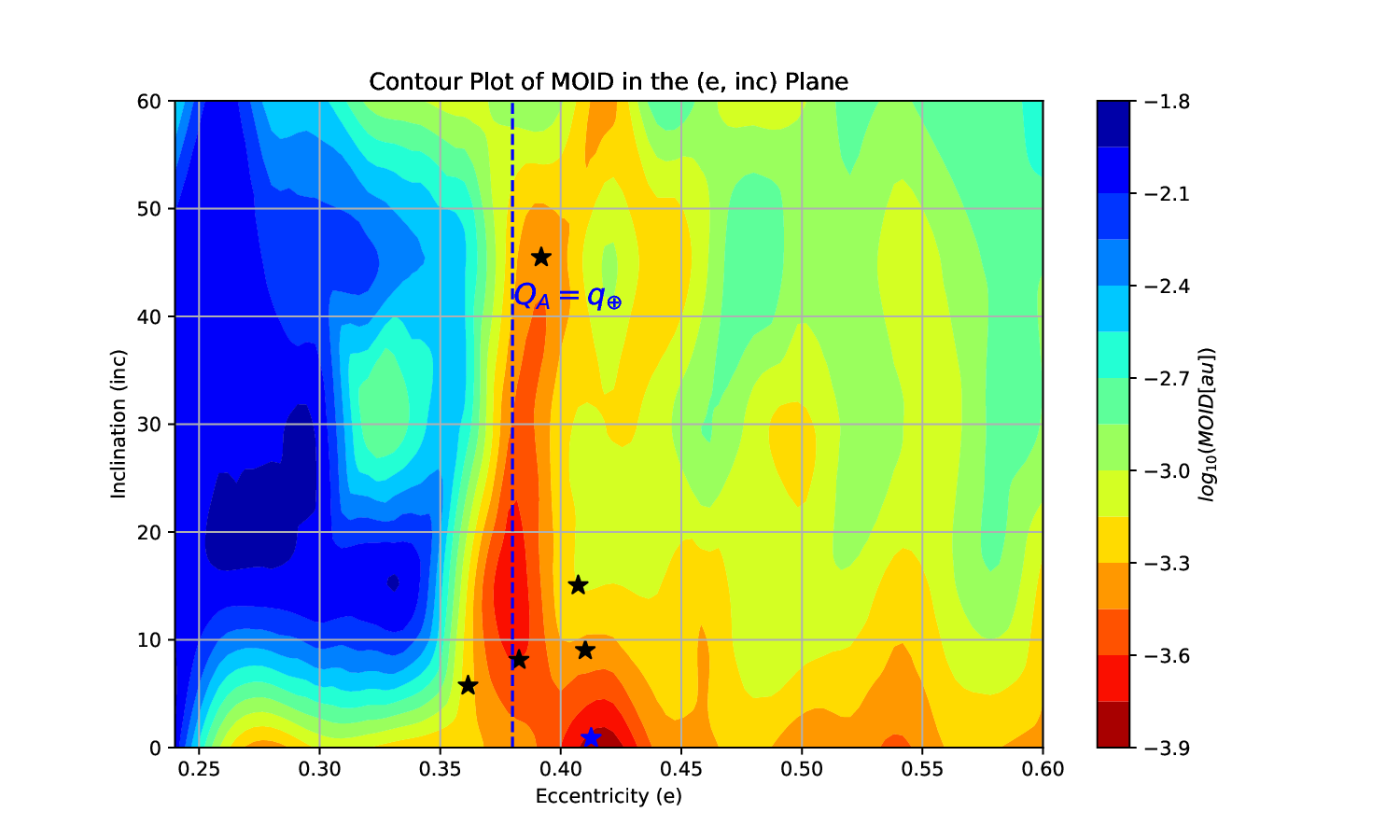}
  \end{minipage}%
  \begin{minipage}[c]{0.45\textwidth}
    \centering \includegraphics[width=2.5in]{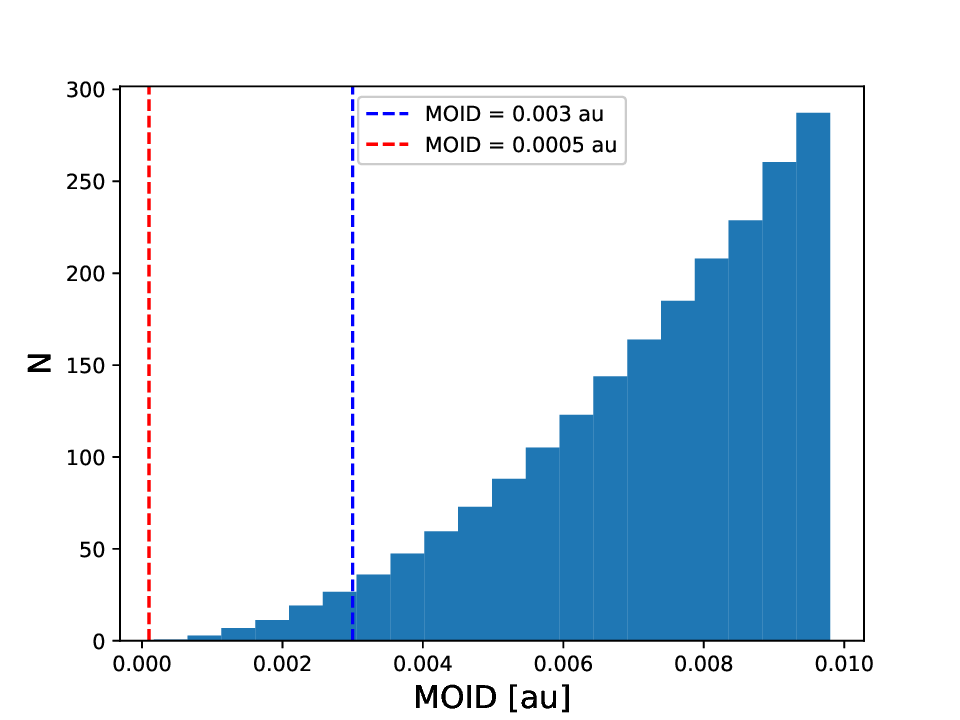}
  \end{minipage}
  \caption{Left panel: MOID contour map with Earth as a function of the initial values of $(e, i)$.  The vertical line shows $e=0.38$, where the Earth's pericenter and the apocenter of the Venus co-orbital asteroid are equal.  Co-orbitals of Venus with a MOID of 0.0005 au or less are shown by the black stars.  The orbital position of 2025 TV$_{10}$ is shown by the blue star. Reproduced with permission from Fig. 7 of \citet{Carruba2025_AandA}.
    Right panel: Histogram of MOID values with Earth for clones of 2025 TV$_{10}$.  The vertical lines display the regions of large (blue line) and high (red line) collision risks.}
  \label{Fig: enc_earth}
\end{figure*}

We then checked where the asteroid is located in the $(e, i)$ MOID contour plane, as recently computed by \citet{Carruba2025_AandA}.  Figure~\ref{Fig: enc_earth}, left panel, shows the orbital location of 2025 TV$_{10}$ in such a domain.  2025 TV$_{10}$ sits very close to the absolute minimum of the MOID, and it is expected to be in one of the most hazardous orbits to Earth among  Venus' known co-orbital asteroids, approaching the Earth within what is called the 
Red Baron scenario, as the Chelyabinsk impactor did \citep{Adamo2013Chelyabinsk}.

This is confirmed by an analysis of the statistics of close encounters with
the Earth obtained from our numerical simulation. Of the simulated clones, 93.3\% were in a sticking orbital configuration for at least part of the first 1000 yr of the simulation, which confirms its likely nature as a co-orbital asteroid of Venus. In addition, following the approach of \citet{Carruba2025_Icarus}, we defined a large collision risk region if the asteroid came closer to Earth than 0.003 au, and a high collision risk region if it approaches at a distance less than 0.0005 au.
If we limit our analysis to timescales of 150 yr, for which the orbits
should still be deterministic, there is an 8.3\% probability that clones of
this asteroid could be reaching the large risk region, while no clone
passed closer than 0.0005 au. Over the full length of the integration, there is a 3.0\% probability of clones of 2025 TV$_{10}$ reaching the large risk region, and a 0.01\% probability of them approaching the high risk region (Fig.~\ref{Fig: enc_earth}, right panel). More remarkable than the statistics of encounters with Earth is the sheer number of close encounters that 2025 TV$_{10}$ may experience in the future.  If we use the simpler approach for setting up initial conditions for clones described in \citet{Carruba2025_AandA} and a longer integration time of 12000 yr, among the known dynamically hazardous asteroids that are engaged in a co-orbital state with Venus, the one with the largest statistics was 524522 Zoozve, with 90298 encounters during the span of the integration. However, 524522 Zoozve (2002 VE$_{68}$) is a quasi-satellite, and its MOID with Earth is currently 0.025887 au. 2025 TV$_{10}$ clones had a total of 167948 encounters, i.e., 85\% more, making it the co-orbital asteroid of Venus with the largest number of close encounters among known dynamically hazardous asteroids.\footnote{On shorter timescales, other co-orbital asteroids of Venus (e.g., 2013 ND$_{15}$) may also experience several close encounters with Earth and have a rather low MOID.  However, on longer timescales, closer to the 12000 yr typically associated with a Venus co-orbital cycle, objects with eccentricities closer to the 0.38 limit for which their apocenter is close to the pericenter of the Earth tend to have larger statistics of encounters and approach Earth at smaller relative velocities.} Because of its high absolute magnitude of 25, this asteroid would likely disintegrate in Earth’s atmosphere if it were to collide (an absolute magnitude of 22 is needed for an asteroid to be classified as a PHA); however, its remarkable dynamical behavior makes it an intriguing asteroid to monitor in the next few years.

\section{Observability from the Vera C. Rubin observatory}
\label{sec: obs}

\begin{figure}[ht!]
  \centering
  \includegraphics[width=2.3in]{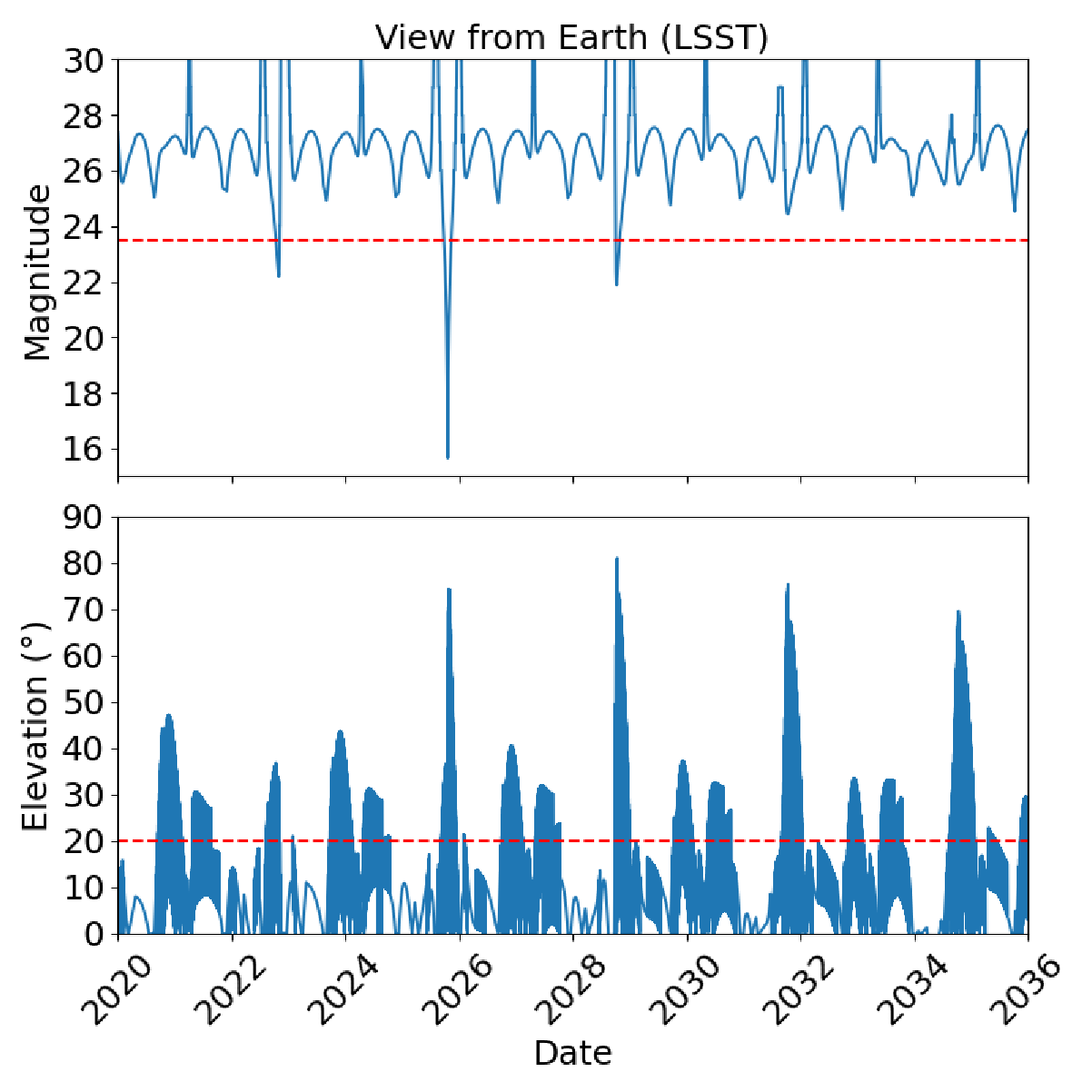}
  \caption{Conditions for 2025 TV$_{10}$ observability from the Rubin Observatory site between 2020 and 2036. The apparent magnitude fluctuations are shown in the upper panel. The Rubin Observatory single-visit detection limit of 23.5 is shown by the horizontal dashed line. The dashed line in the lower panel, which displays the height above the horizon, indicates the $20^\circ$ minimum elevation needed for useful observations.}
  \label{Fig: Rubin}
\end{figure}

Because of its very high absolute magnitude and orbital location, observing
2025 TV$_{10}$ from Earth will be challenging. Using the approach described in \citet{Carruba2025_AandA}, Sect. (5), we computed the apparent magnitude and elevation of this asteroid, as observable from the Vera C. Rubin Observatory in Chile from 2020 to 2036. The ephemerides were also propagated backward to evaluate potential observability. Our results are shown in Fig.~\ref{Fig: Rubin}.
This telescope has a single-visit detection limit of 23.5 magnitudes, and
a $20^\circ$ minimum elevation threshold, which is the threshold for
the ``low-SE twilight survey'' program \citep{Schwamb2023}. Our results show
that the asteroid will be mostly unobservable, with a tiny observability
window of a few days around October 2028.

\section{Conclusions}
\label{sec: concl}

We reported the identification of 2025 TV$_{10}$ as a new co-orbital asteroid of
Venus in a sticking orbit, the 21st known object of this population.
The asteroid is likely to remain a co-orbital asteroid of Venus for timescales of 1200 yr.  More interestingly, its current orbit is the closest one to
the absolute achievable minimum of the MOID with Earth for a known co-orbital of Venus. This asteroid has the largest number of close encounters with Earth, making it the asteroid with the highest collision risk among this population.
Future observations are difficult due to its faint magnitude and limited observability windows, but they are crucial in order to better assess its orbit.

\section{Data availability}

The data and codes produced in this work will be made available upon reasonable request.

\begin{acknowledgements}

  We are grateful for comments from an anonymous reviewer that improved the quality of this work.  We acknowledge the contribution of Dr. Jon Giorgini, who provided the covariance matrix for the elements of 2025 TV$_{10}$ and analyzed the precovery images for the same asteroid. This work made use of the ADAM::Precovery service (B612 Asteroid Institute; https://b612.ai/adam-platform/precovery).

  VC, RS, and OCW acknowledge the support of the Brazilian National Research Council (CNPq, grants 304168/2021-1 (VC), 307400/2025-5 (RS), and 316991/2023-6 (OCW)) and of the São Paulo State Research Support Foundation (Fapesp, grants 2022/11783-5 (OCW) and 2025/01469-0 (VC)).
\end{acknowledgements}

\bibliographystyle{aa} 
\bibliography{mybib}

\begin{appendix}

\section{Appendix: Covariance matrix and orbital elements for 2025 TV$_{10}$}

In the following tables we report the covariance matrix for the orbital solution
of 2025 TV$_{10}$ (Table~\ref{Table: Covariance_matrix}) and the orbital
elements with their uncertainties for the same asteroid
(Table~\ref{Table: Nominal_orbit}).

\begin{table}[H]
\caption{Covariance matrix for the orbital solution of 2025 TV$_{10}$.}
\centering
\begin{tabular}{ccccccc}
\hline
Element & EC & QR [au] & TP [d] & OM [deg]& W [deg] & IN [deg] \\
\hline
EC & 5.432156E-08 & -5.415440E-08 & 6.327905E-06 & -2.048488E-08 & 2.555881E-08 & 2.278728E-09 \\
QR [au] &              & 5.398776E-08 & -6.308433E-06 & 2.042166E-08 & -2.547998E-08 & -2.271715E-09 \\
TP [d] &              &              & 7.371384E-04 & -2.385453E-06 & 2.976540E-06 & 2.654475E-07  \\
OM [deg] &              &              &              & 8.725392E-09 & -1.062963E-08 & -8.599791E-10  \\
W [deg] &               &              &              &              & 1.300815E-08 & 1.072815E-09 \\
IN [deg] &              &              &              &              &              &  9.559140E-11 \\
\hline
\end{tabular}
\tablefoot{EC stands for eccentricity, QR for pericenter, TP for time of passage at pericenter (corresponding to Jun 29.55369 2025, TDB), OM for $\Omega$, W for $\omega$, and IN for inclination.}
\label{Table: Covariance_matrix}
\end{table}

\begin{table}[H]
\caption{Orbital elements and their nominal uncertainties for 2025 TV$_{10}$.}
\centering
\begin{tabular}{lll}
\hline
Elem. & Value & Nominal uncertainties\\
\hline
EC & 0.4125914187681879 & 2.330699E-04 \\
QR & 0.4194460871772284 [au] & 2.323527E-04 [au] \\
TP & 2460856.0536923744 [d] & 2.715029E-02 [d] \\
OM & 294.14463674047386 & 5.351988E-03 [deg] \\
W & 270.71446927420595 [deg] & 6.534771E-03 [deg] \\
IN & 0.88341062385341 [deg] & 5.601857E-04 [deg] \\ 
\hline
\end{tabular}
\label{Table: Nominal_orbit}
\end{table}

\end{appendix}
\label{lastpage}
\end{document}